%% file: main.tex
\documentclass[lettersize,journal]{IEEEtran}
\input{tool}
\begin{document}

\title{NumScout: Unveiling Numerical Defects in Smart Contracts using LLM-Pruning Symbolic Execution}

% \author{Jiachi Chen, Zhenzhe Shao, Shuo Yang, Yiming Shen, Yanlin Wang, Ting Chen, Zhenyu Shan, Zibin Zheng~\IEEEmembership{Fellow,~IEEE}

\author{
    Jiachi~Chen, Zhenzhe~Shao, Shuo~Yang, Yiming~Shen, Yanlin~Wang, Ting~Chen, Zhenyu~Shan, Zibin~Zheng,~\IEEEmembership{Fellow,~IEEE}

    % \thanks{Jiachi Chen is with the Intelligent Transportation and Information Security Laboratory, Hangzhou Normal University, Hangzhou 311121, China, and also with the School of Software Engineering, Sun Yat-sen University, Zhuhai 519082, China (e-mail: \href{mailto:chenjch86@mail.sysu.edu.cn}{chenjch86\allowbreak@mail\allowbreak.sysu\allowbreak.edu.cn})}
    % \thanks{Zhenzhe Shao, Shuo Yang, Yiming Shen, Yanlin Wang, Zibin Zheng are with the School of Software Engineering, Sun Yat-sen University, Zhuhai 519082, China (e-mail:  \href{mailto:shaozhzh3@mail2.sysu.edu.cn}{shaozhzh3\allowbreak@mail2\allowbreak.sysu\allowbreak.edu\allowbreak.cn}; \href{mailto:yangsh233@mail2.sysu.edu.cn}{yangsh233\allowbreak@mail2\allowbreak.sysu\allowbreak.edu\allowbreak.cn}; \href{mailto:shenym7@mail2.sysu.edu.cn}{shenym7\allowbreak@mail2\allowbreak.sysu\allowbreak.edu\allowbreak.cn}; \href{mailto:wangylin36@mail.sysu.edu.cn}{wangylin36\allowbreak@mail\allowbreak.sysu\allowbreak.edu\allowbreak.cn}; \href{mailto:zhzibin@mail.sysu.edu.cn}{zhzibin\allowbreak@mail\allowbreak.sysu\allowbreak.edu\allowbreak.cn})
    % }
    \thanks{Jiachi Chen, Zhenzhe Shao, Shuo Yang, Yiming Shen, Yanlin Wang, Zibin Zheng are with the School of Software Engineering, Sun Yat-sen University, Zhuhai 519082, China (e-mail: \href{mailto:chenjch86@mail.sysu.edu.cn}{chenjch86\allowbreak@mail\allowbreak.sysu\allowbreak.edu.cn}; \href{mailto:shaozhzh3@mail2.sysu.edu.cn}{shaozhzh3\allowbreak@mail2\allowbreak.sysu\allowbreak.edu\allowbreak.cn}; \href{mailto:yangsh233@mail2.sysu.edu.cn}{yangsh233\allowbreak@mail2\allowbreak.sysu\allowbreak.edu\allowbreak.cn}; \href{mailto:shenym7@mail2.sysu.edu.cn}{shenym7\allowbreak@mail2\allowbreak.sysu\allowbreak.edu\allowbreak.cn}; \href{mailto:wangylin36@mail.sysu.edu.cn}{wangylin36\allowbreak@mail\allowbreak.sysu\allowbreak.edu\allowbreak.cn}; \href{mailto:zhzibin@mail.sysu.edu.cn}{zhzibin\allowbreak@mail\allowbreak.sysu\allowbreak.edu\allowbreak.cn})
    }

    \thanks{Ting Chen is with the School of Computer Science and Engineering(School of Cyber Security), University of Electronic Science and Technology of China, Chengdu 611731, China, and also with Kashi Institute of Electronics and Information Industry, Kashi, 844000, China (e-mail: \href{mailto:brokendragon@uestc.edu.cn}{brokendragon\allowbreak@uestc\allowbreak.edu\allowbreak.cn})}

    \thanks{Zhenyu Shan is with the Intelligent Transportation and Information Security Laboratory, Hangzhou Normal University, Hangzhou 311121, China (e-mail: \href{mailto:20100119@hznu.edu.cn}{20100119\allowbreak@hznu\allowbreak.edu\allowbreak.cn})}

    \thanks{Zhenyu Shan is the corresponding author.}
}

\maketitle

\begin{abstract}
In recent years, the Ethereum platform has witnessed a proliferation of smart contracts, accompanied by exponential growth in total value locked (TVL). High-TVL smart contracts often require complex numerical computations, particularly in mathematical financial models used by many decentralized applications (DApps). Improper calculations can introduce numerical defects, posing potential security risks. Existing research primarily focuses on traditional numerical defects like integer overflow, and there is currently a lack of systematic research and effective detection methods targeting new types of numerical defects.
In this paper, we identify five new types of numerical defects through the analysis of 1,199 audit reports by utilizing the open card method. Each defect is defined and illustrated with a code example to highlight its features and potential consequences.
We also propose NumScout, a symbolic execution-based tool designed to detect these five defects. Specifically, the tool combines information from source code and bytecode, analyzing key operations such as comparisons and transfers, to effectively locate defects and report them based on predefined detection patterns. Furthermore, NumScout uses a large language model (LLM) to prune functions which are unrelated to numerical operations. This step allows symbolic execution to quickly enter the target function and improve runtime speed by \improvement.
We run NumScout on \LargeDatasetNum~real-world contracts and evaluated its performance based on manually labeled results. We find that \LargeDefectNum~contracts contained at least one of the five defects, and the tool achieved an overall precision of \OverallPrecision.
\end{abstract}

\begin{IEEEkeywords}
Smart Contracts, Numerical Defects, LLM, Symbolic Execution
\end{IEEEkeywords}

\input{body}

\bibliographystyle{IEEEtran}
\bibliography{ref}

\end{document}

%% file: tool.tex
\usepackage{algorithm}
\usepackage[noend]{algpseudocode}
\usepackage{multirow}
\usepackage{booktabs} % For formal tables
\usepackage{amsmath}
\usepackage{amssymb}
\usepackage{graphicx}
\usepackage{caption}
\usepackage{subcaption}
\usepackage{rotating}
\usepackage{adjustbox}
\usepackage{array}
\usepackage{float}
\usepackage{tabularx}
\usepackage{xcolor}
\usepackage{stfloats}
\usepackage{wrapfig}
\usepackage{listings}
\usepackage{color, colortbl}
\usepackage{balance} % to balance the bibliography
\usepackage{lscape}
\usepackage{xspace}
\usepackage{framed}
\usepackage{syntax}
\usepackage{parcolumns}
\usepackage{pifont}
\usepackage{soul}
\usepackage[tikz]{bclogo}
\usepackage{enumitem}
\usepackage{soul}
\usepackage{makecell}
\usepackage{tcolorbox}
\usepackage{enumitem}
\usepackage{ulem}
\usepackage{diagbox}
\usepackage[hyphens]{url}
\usepackage{hyperref}

\definecolor{verylightgray}{rgb}{.97,.97,.97}
\lstdefinelanguage{Solidity}{
  keywords=[1]{anonymous, assembly, assert, balance, break, call, callcode, case, catch, class, constant, continue, constructor, contract, debugger, default, delegatecall, delete, do, else, emit, event, experimental, export, external, false, finally, for, function, gas, if, implements, import, in, indexed, instanceof, interface, internal, is, length, library, log0, log1, log2, log3, log4, memory, modifier, new, payable, pragma, private, protected, public, pure, push, require, return, returns, revert, selfdestruct, send, solidity, storage, struct, suicide, super, switch, then, this, throw, transfer, true, try, typeof, using, value, view, while, with, addmod, ecrecover, keccak256, mulmod, ripemd160, sha256, sha3}, % generic keywords including crypto operations
  keywordstyle=[1]\color{blue}\bfseries,
  keywords=[2]{address, bool, byte, bytes, bytes1, bytes2, bytes3, bytes4, bytes5, bytes6, bytes7, bytes8, bytes9, bytes10, bytes11, bytes12, bytes13, bytes14, bytes15, bytes16, bytes17, bytes18, bytes19, bytes20, bytes21, bytes22, bytes23, bytes24, bytes25, bytes26, bytes27, bytes28, bytes29, bytes30, bytes31, bytes32, enum, int, int8, int16, int24, int32, int40, int48, int56, int64, int72, int80, int88, int96, int104, int112, int120, int128, int136, int144, int152, int160, int168, int176, int184, int192, int200, int208, int216, int224, int232, int240, int248, int256, mapping, string, uint, uint8, uint16, uint24, uint32, uint40, uint48, uint56, uint64, uint72, uint80, uint88, uint96, uint104, uint112, uint120, uint128, uint136, uint144, uint152, uint160, uint168, uint176, uint184, uint192, uint200, uint208, uint216, uint224, uint232, uint240, uint248, uint256, var, void, ether, finney, szabo, wei, days, hours, minutes, seconds, weeks, years},  % types; money and time units
  keywordstyle=[2]\color{teal}\bfseries,
  keywords=[3]{block, blockhash, coinbase, difficulty, gaslimit, number, timestamp, msg, data, gas, sender, sig, value, now, tx, gasprice, origin},  % environment variables
  keywordstyle=[3]\color{violet}\bfseries,
  identifierstyle=\color{black},
  sensitive=false,
  comment=[l]{//},
  morecomment=[s]{/*}{*/},
  commentstyle=\color{gray}\ttfamily,
  stringstyle=\color{red}\ttfamily,
  morestring=[b]',
  morestring=[b]"
}
\newcommand{\boxmargin}{1mm}
\newtcolorbox{myboxc}{
    colback=gray!15!white,
    arc = 0pt, outer arc = 0pt,
    boxsep=0pt, left = 3pt, right = 0pt, top = 0pt, bottom = 0pt, 
    leftrule=3pt, bottomrule=0pt,toprule=0pt, rightrule=0pt,
    left = \boxmargin, right = \boxmargin, top = \boxmargin, bottom = \boxmargin
}

\newcommand{\MatchReportsNum}{194}
\newcommand{\FilterReportsNum}{109}

\newcommand{\veightperct}{83.4\%}%83.4

\newcommand{\high}{12}
\newcommand{\medium}{33}
\newcommand{\low}{75}

\newcommand{\LargeDatasetNum}{6,617}
\newcommand{\LargeDefectNum}{1,774}

\newcommand{\OverallPrecision}{89.7\%}

\newcommand{\LOC}{1155.9}
\newcommand{\Instrs}{8505.4}
\newcommand{\Funs}{35.5}
\newcommand{\StateVars}{25.4}

\newcommand{\improvement}{28.4\%}

\newcommand{\opensource}{\url{https://github.com/NumScout/NumScout}}

%% file: body.tex
\section{Introduction}
Since the launch of Ethereum~\cite{buterin2014next} in 2015, smart contracts have emerged as a key technology in the blockchain space. Smart contracts are computer programs that automatically enforce predefined agreements on the blockchain, executing transactions without requiring intermediaries. With the rapid development of the Ethereum ecosystem, the number of smart contracts on Ethereum and other blockchain platforms has grown significantly, giving rise to numerous token contracts and decentralized applications(DApps)~\cite{zheng2018blockchain}. Meanwhile, the digital assets involved in these contracts and applications have grown exponentially.

During the design and development of smart contracts, developers frequently handle various numerical computations. In particular, many DApps rely on mathematical financial models that require highly complex computations~\cite{zheng2017overview}. However, owing to the characteristics of the Solidity programming language~\cite{solidity} and the inherent limitations of blockchain platforms, smart contracts are susceptible to various numerical defects. In this paper, we define numerical defects as all numerical-related \textit{errors, vulnerabilities, or flaws} that can lead to unexpected outcomes or deviate from the original code’s intent~\cite{lyu1996handbook}. Notably, numerical defects involve not only security issues but also design flaws, which can increase the long-term risk of the smart contracts.

%Since these contracts directly manage significant amounts of digital assets and cannot be modified once deployed, ensuring their numerical security before deployment is critical. The contract defects refer to errors, vulnerabilities, or flaw that can lead to unexpected outcomes or deviate from the original code’s intent~\cite{lyu1996handbook}. These defects involve not only security issues but also design flaws, which can increase the long-term risk of the contract.

Numerous real-world hacking incidents caused by numerical defects have already resulted in severe financial losses for both project teams and users. Although common numerical security defects, such as integer overflow and type conversion errors~\cite{arithmetic-issue}, have been identified and mitigated through solutions like the SafeMath library~\cite{safemath} and the introduction of new security mechanisms in Solidity v0.8~\cite{sol08changes}, new types of numerical defects continue to emerge in practice. For example, over \$2.12 million in assets were stolen from \textit{Balancer}~\cite{balancer} due to a precision-related issue~\cite{balancerloss}.
These numerical defects pose significant threats to the security and reliability of contracts. However, a systematic study that classifies new types of numerical defects and provides corresponding detection methods and tools is still lacking. 

To fill the gap, we first conducted an empirical study to define new types of numerical defects by analyzing 1,199 audit reports using an open card sorting method~\cite{spencer2009card}. Based on this analysis, we identified five categories of new numerical defects, i.e., \textit{Div In Path}, \textit{Operator Order Issue}, \textit{Minor Amount Retention}, \textit{Exchange Problem}, and \textit{Precision Loss Trend}. We present examples for each defect type and propose corresponding mitigation strategies to enhance the quality and robustness of smart contracts.

\IEEEpubidadjcol

Then, we developed a tool named NumScout, designed to detect the five new types of numerical defects in real-world contracts. NumScout leverages the reasoning capabilities of Large Language Models (LLMs)~\cite{wei2022emergent} and combines source code level information with bytecode analysis to enhance detection efficiency in complex contracts. Specifically, NumScout first uses LLM-based pruning to exclude functions unrelated to numerical operations or token transfers. This step is designed to mitigate the path explosion problem in symbolic execution and accelerate the analysis process of the tool. Due to the complex semantics and call relationships of contracts, static pruning methods based on simple rule matching fail to meet the requirements. LLMs can perform reasoning and analysis at the high-level semantic layer and across multi-level calls. By leveraging a multi-role collaboration strategy, they reduce response randomness and error, thereby accomplishing the pruning task effectively. Then, based on predefined patterns and a symbolic execution framework, the tool performs symbolic execution at the bytecode level, incorporating features from the source code for further analysis. It focuses on key operations such as comparisons and transfers, and identifies defects through various methods, including constructing and analyzing expression operator order trees, extracting comparison statements from bytecode, and analyzing token flows.

To demonstrate the prevalence of the five defined numerical defects and evaluate the efficacy of NumScout, we filter \LargeDatasetNum~real-world smart contracts which are frequently used by users on Ethereum~\cite{smartcontractsanctuary}, ensuring that the contracts in our experimental dataset have actual value rather than toy contracts. We apply NumScout to these \LargeDatasetNum~smart contracts and find that \LargeDefectNum~contracts contain at least one of the five defined defects. Then, we randomly sample contracts with a 95\% confidence level and a 10\% confidence interval for manual labeling. The results show that the tool achieves an overall precision of \OverallPrecision. In addition, we conduct ablation experiments to verify the effectiveness of GPT-based pruning. The experiments demonstrate that pruning enables symbolic execution to quickly enter the target functions, improving runtime speed by \improvement~and detecting more defects.

The main contributions of our work are as follows:

\begin{itemize}[left=10pt]
    \item We summarize and define five new types of numerical defects based on analyzing 1,199 audit reports. For each defect, we provide its definition with a code example for better illustration. Furthermore, we outline possible solutions to enhance development security.
    \item We develop NumScout, the first tool designed for the defined numerical defects. NumScout employs LLM pruning functions and recovers source-level features from bytecode during symbolic execution to identify designed defect patterns more efficiently. %Additionally, NumScout is an extensible framework that supports all versions of the Solidity compiler.
    \item We evaluate NumScout’s performance on \LargeDatasetNum~real-world smart contracts and discover that \LargeDefectNum~contracts contain at least one defined defect. Moreover, in a manually labeled dataset created through random sampling, our approach achieves an overall precision of \OverallPrecision.
    \item We make the source code of NumScout, all experimental data, and analysis results publicly available, along with detailed Markdown files at \opensource.
\end{itemize}

%The structure of this paper is as follows: Section \ref{section:backgrond} introduces the background knowledge of smart contracts, numerical defects, and LLM. Section \ref{section:defects} defines and presents examples of 5 new types of numerical defects. Section \ref{section:methodology} describes the architecture and key technologies of NumScout. Section \ref{section:experiment} presents the experiments conducted to evaluate NumScout. Section \ref{section:discussion} discusses the experimental results, tool design, and methods for addressing defects. Section \ref{section:relatedwork} reviews related work, and Section \ref{section:conclusion} provides the conclusion.

\section{Background} \label{section:backgrond}

\subsection{Numerical Operations in Solidity and Integer Overflow}

Solidity is the most popular programming language for smart contracts on Ethereum. The computations in Solidity smart contracts are performed using arithmetic opcodes, e.g., \textit{ADD} and \textit{MUL}~\cite{evmcodes}. Due to the inherent characteristics of the language and the limitations of the blockchain platform, for example, maintaining the consistency of the public ledger and reducing computational resource consumption, Solidity only supports integers and does not support floating-point numbers, which can introduce certain numerical issues. In traditional numerical detection, integer overflow is one of the most common defects in smart contracts~\cite{atzei2017survey,wang2021ethereum,khan2020survey,samreen2021survey}. An integer overflow defect occurs when the result of an arithmetic operation exceeds the range of its data type, producing an outcome that deviates from expectations. Since smart contracts typically use integers to represent asset amounts and other numerical values, calculations involving these numbers may experience overflow or underflow under malicious input from attackers, resulting in asset loss. Several notable attacks have occurred due to this defect, including \textit{BeautyChain} token (BEC)~\cite{bec} attack, \textit{SmartMesh} token (SMT)~\cite{smt} attack, and \textit{UselessEthereumToken} token (UET)~\cite{uet} attack.

The developer community has built security libraries to prevent overflows, such as the widely adopted SafeMath library~\cite{safemath}, developed by the well-known blockchain security team OpenZeppelin~\cite{openzeppelin}, which ensures the correctness of calculation through boundary checks. Starting from version v0.8.0, the Solidity compiler introduces arithmetic checking mechanisms~\cite{sol08changes}, which embed overflow detection into the compiled bytecode. If an overflow occurs during a transaction, the EVM~\cite{buterin2014next} will throw an error and revert. However, although traditional integer overflows have been largely mitigated, increasingly complex contract scenarios are giving rise to new types of numerical defects that are easily overlooked.

\subsection{Smart Contract Audit Report}

Smart contract auditing is an important process in the blockchain ecosystem, focusing on identifying vulnerabilities and defects in smart contract code. Auditors from professional auditing teams assess the code to identify potential defects, ensuring that the contract operates as intended and adheres to best practices. 
Audit reports provide a comprehensive analysis of smart contracts, detailing all identified defects and their impacts, assigning severity levels, and offering recommended remediation strategies. These reports serve as essential documentation for developers, investors, and users, enhancing the transparency and trustworthiness of the project’s contracts. Given the irreversibility of blockchain transactions, thorough auditing is vital to prevent financial losses and maintain the integrity of DApps.

\subsection{Large Language Models}

Large Language Models (LLMs)~\cite{zhao2023survey,chang2024survey} are deep learning-based natural language processing models that possess powerful language understanding and generation capabilities. The GPT (Generative Pre-trained Transformer) series, developed by OpenAI~\cite{openai}, is a prominent representative of LLMs. GPT utilizes the Transformer~\cite{vaswani2017attention} architecture and is trained on extensive corpora, including source code descriptions of various programming languages and known defects. With this knowledge, GPT can understand and interpret source code, enabling zero-shot learning~\cite{kojima2022large}. The latest version, GPT-4o~\cite{gpt-4o}, supports a 128k context length, making it suitable for complex and multi-step tasks. While LLMs and GPT have shown significant potential in fields such as smart contract analysis, trustworthiness and accuracy remain critical research challenges~\cite{chen2023chatgpt,chen2024identifying}.

Multiple studies have demonstrated that LLMs exhibit excellent code understanding capabilities. They have great ability in understanding code syntax and semantics, including Abstract Syntax Tree (AST) and Control Flow Graph (CFG)~\cite{ma2024unveiling}. LLMs have been applied in multiple fields that require code understanding~\cite{zheng2025towards}. For example, they are used for analyzing inconsistencies in code comments~\cite{nam2024using}. They also serve as the foundation for developer assistance tools~\cite{zhang2024detecting}. Furthermore, in the field of smart contract vulnerability detection, LLMs act as code understanding tools to identify logical vulnerabilities~\cite{sun2024gptscan,ma2024combining,ding2025smartguard,wang2024contracttinker,wu2024advscanner}.

\label{LLMscapabilities}

\subsection{Symbolic Execution}

%The defect detection technology based on symbolic execution for smart contracts has the main idea of symbolizing the storage variables and external inputs in the contract. 
Symbolic execution-based defect detection for smart contracts primarily involves symbolizing the storage variables and external inputs within the contract.
Smart contracts are typically executed on the Ethereum Virtual Machine (EVM). The EVM features a stack-based architecture and is responsible for interpreting and executing the opcodes of contracts. To describe the execution flow of contracts more clearly, the Control Flow Graph (CFG) is often utilized. The CFG represents the program's basic blocks and their control flow relationships, aiding the analysis of the reachability of different execution paths.
During the symbolic execution process, a set of path constraints is maintained for each explored execution paths. These constraints consist of conditions related to symbolic variables, which describe the current execution state of the contract. The satisfiability modulo theories (SMT) solver is used to evaluate these constraints and determine whether specific conditions are satisfied, such as identifying inputs that may trigger vulnerabilities or verifying the solvability of constraints after adding new conditions. The typical workflow of symbolic execution tools is as follows: Execute the program's opcodes sequentially, symbolize variables and external inputs as they are encountered, update the program context state and add new conditions to path constraints during the process. While exploring all the executable paths of the program, the tool assesses the satisfiability of security-related conditions to detect potential security issues.

Traditional symbolic execution methods often encounter the path explosion problem, which can prevent the completion of detection within a reasonable timeframe. To address this issue, pruning methods are necessary to mitigate path explosion and accelerate the analysis process. As the complexity of smart contracts has increased in recent years, particularly in terms of semantics and call relationships, traditional pattern-based pruning methods %do not perform effectively.
tend to be less effective.
% face challenges in achieving both accuracy and efficiency.
In contrast, LLMs can recognize high-level semantics and multi-level calls, making them well-suited for completing the pruning tasks of complex contracts.

\section{New Numerical Defects} \label{section:defects}

In this section, we explain how the five new types of numerical defects are identified and provide definitions and examples for each defect.

\subsection{Data Source}

To identify and define new types of numerical defects, we analyze 1,199 audit reports collected by DAppScan~\cite{zheng2024dappscan}. %DAppScan is a public dataset containing smart contracts from real-world DApp projects, along with their corresponding audit reports. These reports are collected from the official websites, social media, or Web3 sites of multiple well-known blockchain security teams, such as BlockSec~\cite{blocksec} and PeckShield~\cite{peckshield}. The author team of DAppScan recruits 22 participants and spends 44 person-months analyzing all the audit reports to build a large-scale SWC (Smart Contract Weakness Classification) ~\cite{swc} dataset. These audit reports serve as a rich resource, with many non-SWC defects still awaiting further identification. By matching keywords such as ``precision loss'', ``rounding error'' and similar terms, we extract \MatchReportsNum~audit reports related to numerical defects from the DAppScan dataset for further analysis.
DAppScan is a public dataset containing audit reports collected from the official websites, social media, and Web3 sites of 29 well-known blockchain security teams, such as Openzeppelin~\cite{openzeppelin} and Consensys~\cite{consensys}. These audit reports serve as a rich resource, revealing numerous numerical defects found in real-world projects. %By matching keywords ``precision loss'' and ``rounding'', we extract \MatchReportsNum~audit reports related to numerical defects from the DAppScan dataset for further analysis.
We adopt a keyword matching approach to filter reports content related to numerical defects, while employing Snowball Sampling~\cite{goodman1961snowball} strategy to ensure the completeness of the keyword list. Initially, we filter the audit reports by matching the keywords ``precision'' and ``rounding''. During the review of the report content, we record new keywords related to numerical defects and add them to the keyword list for filtering new reports. Ultimately, we filter a total of \MatchReportsNum~audit reports using 25 keywords for further analysis. For the complete keyword list, please refer to our online repository.

\subsection{Audit Report Analysis}

\begin{figure}[b]
% \vspace{-1em}
    \centering
    \includegraphics[width=\linewidth]{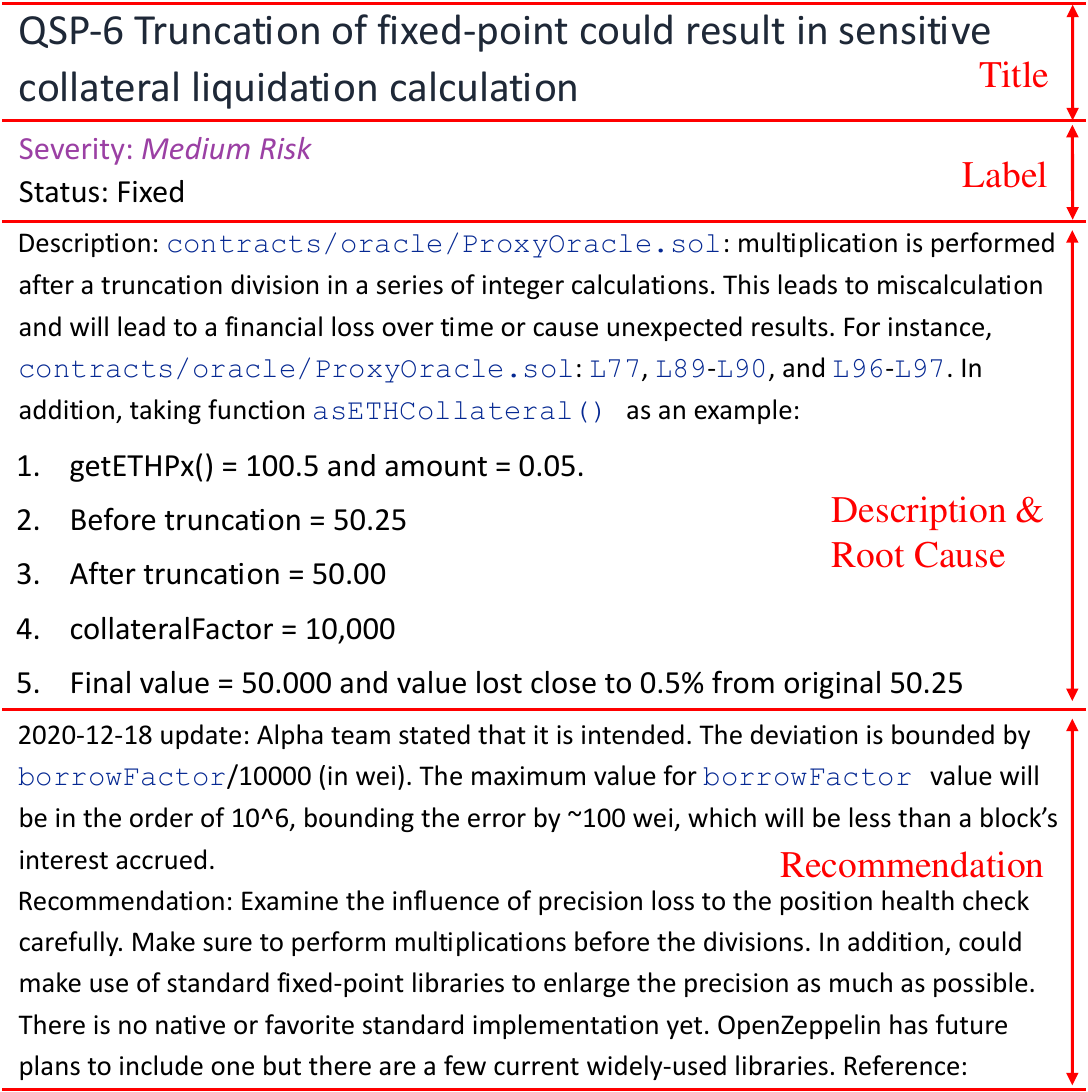}
    % \vspace{-1em}
    \caption{Example of a card of audit reports}
   % \vspace{-1em}
    \label{fig:opencard}
\end{figure}

\subsubsection{Manual Filtering} In the previous subsection, we describe the collection of \MatchReportsNum~publicly available audit reports from renowned blockchain security teams. However, some of these reports are not directly related to numerical defects. For example, certain reports mention ``precision loss'' but only discuss its risk or offer general advice to users, instead of detailing specific defects in the code. Therefore, we manually remove reports that lack specific defect descriptions. After filtering, we find \FilterReportsNum~reports directly related to numerical defects from the initial set of \MatchReportsNum~security reports.
%It is worth noting that we did not discard the removed audit reports. Instead, we extracted the potential threats and useful mitigation strategies mentioned within them.

\subsubsection{Open Card Sorting} To ensure accuracy, we use the open card sorting ~\cite{spencer2009card} approach to analyze and categorize the filtered audit reports related to numerical defects. In this process, we consider two aspects to ensure the representativeness and significance of the defects, i.e., the reproducibility of the code issue and severity as assessed by the security teams. Some issues may be tightly coupled with specific applications and not reproducible; we do not classify these as representative defects. Additionally, we focus on the labels assigned by security teams in the reports to assess the severity of the identified defects.

For each numerical defect mentioned in the audit reports, we create a card comprising four sections to organize the content.
Following the detailed steps outlined in ~\cite{chen2020defining}, we begin by randomly selecting 40\% of the cards for the first round of classification. First, we read the titles and descriptions of the reports to understand the relevant defects. Next, we inspect the problematic code to identify the root cause and cross-reference it with the audit reports. Finally, we review the recommended solutions suggested by the security teams to understand how to address the defects and record the severity level assigned by the team.

In the second round of classification, two authors independently categorize the remaining 60\% of the cards following the same steps described in the first round. We then compare their results and discuss discrepancies. Next, we remove uncommon defects and ultimately classify the remaining defects into five types. Among the classified reports, \high~are labeled as high, \medium~as medium, and \low~as low severity.

Figure \ref{fig:opencard} shows a card example of an audit report describing a numerical defect. The card contains a title, assigned label, description, root cause, and recommended solution. From the report, we learn that the contract contains a defect where division is incorrectly performed before multiplication. We then locate the referenced code (i.e., contracts/oracle/ProxyOracle.sol: L77, L89-L90, and L96-97) to further confirm the root cause and verify the presence of this defect. The report also provides an example of miscalculations caused by this defect, demonstrating its exploitability and the potential consequences. Due to the reproducibility of this defect and its frequent occurrence in audit reports, we classify it as a distinct defect type named ``\textit{Operator Order Issue}''.

\subsection{Defects Definition}
Based on the analysis of the audit reports mentioned in the previous section, we have summarized five new types of numerical defects. Table \ref{table:Definition} provides a brief definition of each defect, followed by a detailed definition and code example for each defect pattern.

\begin{table}[h]
  \centering
  \normalsize
  \setlength\tabcolsep{1.5pt}
  \caption{Definitions of the Five Defects}
\resizebox{\columnwidth}{!}{
\begin{tabular}{>{\centering\arraybackslash}m{0.3\linewidth} | >{\arraybackslash}m{0.69\linewidth}}
    \hline
    \multicolumn{1}{c|}{\textbf{Contract Defect}} & \multicolumn{1}{c}{\textbf{Definition}} \\
    \hline
    \hline
    \textit{Div In Path} & The use of division in comparison conditions affects the execution path. \\
    \hline
    \textit{Operator Order Issue} & Dividing before multiplying amplifies precision loss. \\
    \hline
    \textit{Minor Amount Retention} & When multiple parties share profits, indivisible amounts remain trapped in the contract and cannot be withdrawn. \\ 
    \hline
    \textit{Exchange Problem} & Errors in token amount calculations during token exchanges create rounding issues or profit opportunities. \\
    \hline
    \textit{Precision Loss Trend} & Incorrect rounding methods lead to unreasonable allocation of precision loss. \\
    \hline
\end{tabular}}
 \label{table:Definition}
% \vspace{-1em}
\end{table}

\noindent{\bfseries (1) Div In Path.} The Solidity programming language does not support floating-point numbers, so all division operations result in integer division~\cite{soliditydivision}. When the result is not a whole number, only the integer part is retained, leading to precision loss. If division is used within a conditional statement, this inherent precision loss can potentially alter the program’s execution path and cause unexpected results. Consequently, users may be misled by this defect and pass incorrect values. The blockchain security team ChainSecurity~\cite{chainsecurity} issues a warning about this defect in their audit report for the \textit{Angle Protocol Borrowing Module} project.
%For example, in a conditional statement like \verb|if(x/100>3)|, users who are not familiar with Solidity might assume that as long as $x>3*100$, the condition will be met, and the program will enter the expected execution branch. However, in reality, for values of $x$ between 300 and 400, the condition is not satisfied, causing the program to take an unintended path.
\label{ifGT}

{\bfseries Example:} As shown in Figure \ref{fig:DIP}, users can purchase tokens by sending ether through the $getTokens$ function. The internal conditional check restricts the minimum amount of ether, with the $minAmount$ set to 3. However, users who are not familiar with Solidity may assume that sending more than $3\;ether$ will satisfy the condition, and the program will enter the expected execution branch for token purchase. In reality, the requirement is that $msg.value$ must exceed $4\;ether$. For amount between $3\;ether$ and $4\;ether$, e.g., $3.5\;ether/1\;ether=3$ instead of 3.5, this condition is still not met, preventing users from buying tokens. If the contract does not handle such situations, users may lose their funds without receiving any tokens. Malicious contracts can exploit this defect to scam inexperienced users.

\begin{figure}[h]
\centering
% \vspace{-1em}
\begin{lstlisting}
function getTokens(address _to, uint256 _amount) public payable returns (bool) {
    if(msg.value / 1 ether > minAmount) {
        /* buy tokens */}}
\end{lstlisting}
% \vspace{-0.5em}
\caption{An example of Div In Path defect.}
% \vspace{-0.5em}
\label{fig:DIP}
\end{figure}

\noindent{\bfseries (2) Operator Order Issue.} The most common defect regarding calculation order is performing division before multiplication. This defect leads to incorrect calculation results because multiplication can amplify the precision loss introduced by division. Therefore, in programming practices, when both multiplication and division appear in an expression, it is generally recommended to perform multiplication first and then division to minimize precision loss. However, in today's increasingly complex contracts, developers often overlook this principle, and cases where division is done before multiplication frequently occur. \textit{Operator Order Issue} is also the most common defect in audit reports. The security team QuillAudits~\cite{quillaudits} includes a warning about this defect in their audit report for the \textit{Alium Finance Smart Contract} project. 

{\bfseries Example:} %The security team QuillAudits~\cite{quillaudits} includes a warning about this defect in their audit report for the \textit{Alium Finance Smart Contract} project. 
Figure \ref{fig:OOI} shows the $updatePool$ function, which handles the logic for retrieving staking rewards $almReward$, with 10\% of the rewards allocated to the developer, i.e., $devReward$. %It updates the reward per share for the specified liquidity pool and ensures that the reward calculation for the pool stays consistent with the current block height on the blockchain.
This calculation results in a precision loss of one decimal. If $almReward=199$, then $devReward=(199/100)*10=10$. However, the calculation of 10\% of $almReward$ should result in 19. If the code {\normalsize \texttt{devReward = almReward.mul(10).div(100)}} is used instead, then the result will be correct. This defect can lead to financial losses for developers over time. 

\begin{figure}[h]
\centering
% \vspace{-1em}
\begin{lstlisting}
function updatePool(uint256 _pid) public {
    // deduct 10% for the developers
    uint256 devReward = almReward.div(100).mul(10);
    _safeAlmTransfer(devaddr, devReward);}
\end{lstlisting}
% \vspace{-0.5em}
\caption{An example of Operator Order Issue defect.}
% \vspace{-0.5em}
\label{fig:OOI}
\end{figure}

%This calculation results in a loss of precision beyond the first decimal place. e.g. if $almReward=199$ then $devReward=(199/100)*10=10$. If the code recommended by QuillAudits, \lstinline|devReward = almReward.mul(10).div(100)|, is used instead, then $devReward=(199*10)/100=19$. This defect can lead to financial losses for developers over time. In other cases, it may also result in unexpected outcomes. For example, Quantstamp’s report on \textit{Alpha Homora V2} and PeckShield’s report on \textit{VeChain} mention that the ``division-before-multiplication'' defect results in financial calculation errors.

\noindent{\bfseries (3) Minor Amount Retention.} This defect typically arises in scenarios where multiple participants share rewards or withdraw funds. On blockchain platforms, numerous game contracts involve players investing funds to participate, with winners dividing the rewards. During the distribution of rewards, if the total amount is not divisible by the number of users, a small portion of the funds will remain stuck in the contract, unable to be withdrawn.
If the withdrawn tokens are tied to a liquidity pool, leftover tokens could affect the ratio, leading to economic losses. The security team Dedaub~\cite{dedaub} issues a warning about this defect in their audit report for the \textit{GoodGhosting} project.

{\bfseries Example:} %The security team Dedaub~\cite{dedaub} issues a warning about this defect in their audit report for the \textit{GoodGhosting} project. The \textit{GoodGhosting} contract implements an investment game that incentivizes players to maintain an investment plan. In Figure \ref{fig:MAR}, players can withdraw their funds and claim interest rewards through the $withdraw$ function. These interest rewards are evenly distributed among all winning players.
%Dedaub pointed out that the $totalGameInterest$ may not be divisible by $winners.length$, resulting in a minor amount of funds remaining in the contract and being unable to be withdrawn. It is worth noting that the consequences of this defect are not limited to the inability to withdraw small amounts; if the tokens involved are tied to a liquidity pool, improper withdrawal could affect the ratio, leading to economic losses.
The code shown in Figure \ref{fig:MAR} comes from an investment game contract that incentivizes players to participate in the game and maintain the investment plans. Player can withdraw their funds and claim interest rewards generated within the game through the $withdraw$ function. These interest rewards are evenly distributed among all winning players. There is a defect where the $totalGameInterest$ may not be divisible by $winners.length$, resulting in a minor amount of funds remaining in the contract and being unable to be withdrawn. The retained amount can affect the contract's state, making actions that reference that state unsafe. Specifically, if there are other contracts associated with the $daiToken$ balance of this investment pool, such as trading pairs composed of $daiToken$ and other tokens, the retained amount can impact the ratio between the two, leading to financial security issues.

\begin{figure}[h]
\centering
% \vspace{-1em}
\begin{lstlisting}
function withdraw() external virtual {
    // calc interest reward shared by all winners
    payout = payout.add(totalGameInterest.div(winners.length));
    require(IERC20(daiToken).transfer(msg.sender, payout),"Fail to transfer");}
\end{lstlisting}
% \vspace{-0.5em}
\caption{An example of Minor Amount Retention defect.}
% \vspace{-0.5em}
\label{fig:MAR}
\end{figure}

\noindent{\bfseries (4) Exchange Problem.} Token exchanges are fundamental in various scenarios, such as purchasing tokens, providing liquidity, and trading tokens. If the numerical operations involved in the exchange process are not handled properly, issues may arise, including exchange rounding and zero-cost profit opportunities. The former results in users losing their input tokens while receiving zero output tokens. The latter allows users to obtain output tokens without providing any input tokens. The security team Trail of Bits~\cite{trailofbits} reports this defect in their audit of the \textit{Balancer Finance} project.

{\bfseries Example:} %The security team Trail of Bits~\cite{trailofbits} reports this defect in their audit of the \textit{Balancer Finance} project.
As shown in Figure \ref{fig:EP}, the function $joinPool$ allows users to inject assets into the liquidity pool and receive corresponding pool shares (which is also an ERC20~\cite{erc20} token, referred to as pool token). Therefore, there is a token exchange process involved here. The user inputs $poolAmountOut$ to indicate the amount of pool tokens they want to receive. Internally, the function calculates the exchange ratio based on the desired amount of pool tokens and the total amount of pool tokens. It then calculates the number of liquidity tokens that the user needs to contribute based on the current total of liquidity tokens held by the pool. During this process, users may receive pool tokens without having to contribute any liquidity tokens. The calculation method for $bmul$ is as follows $c=\frac{(a*b)+\frac{BONE}{2}}{BONE}$. Thus, the final expression for $tokenAmountIn$ is:

% \vspace{-1em}
$$tokenAmountIn = \frac{bal*\frac{poolAmountOut}{poolTotal}+\frac{BONE}{2}}{BONE}$$

BONE is set to $10^{18}$. Suppose the condition $bal*\frac{poolAmountOut}{poolTotal}<5*10^{17}$ is satisfied, a quantity of $poolAmountOut$ pool tokens will be generated, while the user contributes no liquidity tokens, resulting in $tokenAmountIn = 0$.
This situation occurs if the token has low liquidity or has decimals precision lower than 18, e.g., USDT~\cite{usdt}, USDC~\cite{usdc}, and XRP~\cite{xrp}, which all hold high market values, have only 6 decimals. According to data from Etherscan, these three tokens are all ranked in the top 5 by market capitalization~\cite{toptoken}.

\begin{figure}[h]
\centering
% \vspace{-1em}
\begin{lstlisting}
function joinPool(uint poolAmountOut, uint[] calldata maxAmountsIn) external _logs_ _lock_ {
    // calc swap ratio with input and poolTotal
    uint poolTotal = totalSupply();
    uint ratio = bdiv(poolAmountOut, poolTotal);
    uint bal = _records[t].balance;
    // calc amount user should contribute with ratio
    uint tokenAmountIn = bmul(ratio, bal);}
\end{lstlisting}
% \vspace{-0.5em}
\caption{An example of Exchange Problem defect.}
% \vspace{-0.5em}
\label{fig:EP}
\end{figure}

\noindent{\bfseries (5) Precision Loss Trend} There are three rounding methods for division: rounding down, rounding to the nearest integer, and rounding up. In Solidity, the default behavior for division is rounding down, which returns the largest integer less than or equal to the exact division result, denoted as $floor(x)$. Rounding to the nearest integer adjusts the division result up or down based on the decimal place, and in formulas, it is expressed by adding $(denominator/2)$ to the $numerator$, denoted as $round(x)$. Rounding up returns the smallest integer greater than or equal to the normal division result, expressed as adding $(denominator-1)$ to the $numerator$, denoted as $ceil(x)$. Using different rounding methods can cause different tendencies in the calculation results, and incorrect tendencies can lead to unexpected consequences. The security team PeckShield's~\cite{peckshield} audit report on \textit{OneSwap} includes this defect.

{\bfseries Example:} %PeckShield's~\cite{peckshield} audit report on \textit{OneSwap} includes this defect. In Figure \ref{fig:PLT}, the function $\_dealWithPoolAndCollectFee$ is responsible for handling transactions within the trading pool and collecting fees. The core functionality of the function is to update the reserves of the trading pool based on the given context parameters $ctx$, while also calculating the amount of tokens that should be paid to the user, i.e., the transaction sender.
In Figure \ref{fig:PLT}, the function $\_dealWithPoolAnd\allowbreak CollectFee$ is responsible for handling transactions within the trading pool and collecting fees. 
At this point, the fee calculation uses the standard $floor(x)$ method, rounding the result down. The tokens amounts user receives is the total amount minus the fee, and user obtains the small fractional amount discarded during the rounding down process. This means the calculation tends to favor users, allowing users to receive more tokens. However, in AMM-based DEX scenarios~\cite{uniswap}, the calculation should favor the liquidity pool to protect the interests of liquidity providers. Therefore, the fee should be rounded up, ensuring that more tokens remain in the liquidity pool. Use {\normalsize \texttt{fee = (amountToTaker * feeBPS + 9999) / 10000} to replace the original code.

\begin{figure}[h]
\centering
% \vspace{-1em}
\begin{lstlisting}
function _dealWithPoolAndCollectFee(Context memory ctx, bool isBuy) internal returns (uint) {
    // calc transaction fee
    uint fee = amountToTaker * feeBPS / 10000;
    // calc amount user gets after deducting fee
    amountToTaker -= fee;
    _transferToken(token, ctx.order.sender, amountToTaker, ctx.isLastSwap);
    return amountToTaker;}
\end{lstlisting}
% \vspace{-0.5em}
\caption{An example of Precision Loss Trend defect.}
% \vspace{-0.5em}
\label{fig:PLT}
\end{figure}

% Another scenario affected by this defect is unfair distribution of benefits. For example, when multiple participants split profits proportionally, as shown in the Figure \ref{fig:PLT2} below. At this point, the calculation tends to favor $BB$, resulting in a rounded-up value, while $AA$ corresponds the rounded-down result. Consequently, $BB$ holders receive more. It is crucial to carefully consider the tendency of precision loss and apply the most suitable rounding method for different scenarios.
Another scenario affected by this defect is unfair distribution of benefits as shown in the Figure \ref{fig:PLT2} below. At this point, $BB$ receives a rounded-up value, while $AA$ corresponds the rounded-down value. It is crucial to carefully consider the tendency of precision loss and apply the most suitable rounding method for different scenarios.

\begin{figure}[h]
\centering
% \vspace{-1em}
\begin{lstlisting}
function _updatePrices() internal {
    // calc AA's earnings based on ratio
    AAGain = gain * trancheAPRSplitRatio / FULL_ALLOC;
    // sub AA's earnings from total to obtain BB's
    BBGain = gain - AAGain;}
\end{lstlisting}
% \vspace{-0.5em}
\caption{Another example of Precision Loss Trend defect.}
% \vspace{-0.5em}
\label{fig:PLT2}
\end{figure}

\section{Methodology} \label{section:methodology}
In this section, we introduce the methods for detecting the aforementioned defects. We first provide an overview of our approach, followed by detailed explanations of two main components: GPT-based pruning and symbolic execution. For the latter, we further elaborate on instruction-level details and operational features.

\begin{figure*}[t]
% \vspace{-1em}
    \centering
    \includegraphics[width=\linewidth]{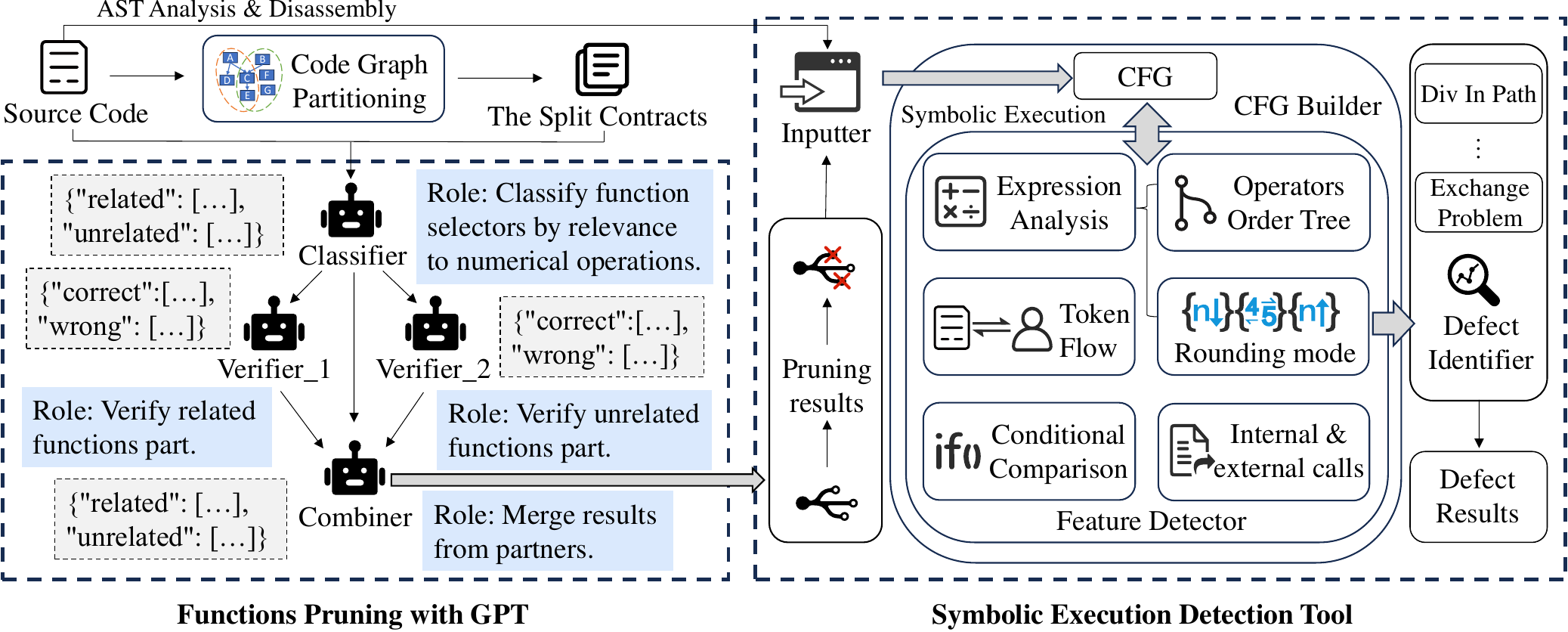}
    % \vspace{-1em}
    \caption{An overview of the approach of NumScout.}
    % \vspace{-1em}
    \label{fig:workflow}
\end{figure*}

\subsection{Overview}
Figure \ref{fig:workflow} presents an overview of NumScout. NumScout consists of two main components: the GPT-based function pruning and the symbolic execution detection tool. Specifically, users provide Solidity source code as input. If the code is too lengthy and exceeds the input limit of GPT, we perform subgraph segmentation. This involves analyzing the contract's abstract syntax tree (AST) and constructing a call graph starting from each entry function based on the internal call relationships. The call graph is then used to break the large contract into several smaller contracts, each with a complete function call chain. The segmented code is subsequently passed to GPT for pruning.

The pruning component of GPT involves four roles: a Classifier, two Verifiers, and a Combiner, which work collaboratively to enhance the accuracy of pruning. The Classifier generates preliminary relevance judgments based on whether the functions in the contract involve numerical operations or fund transfers, classifying them as either ``related'' or ``unrelated'' and sending the results to the two Verifiers. Each Verifier independently verifies a specific subset of the classification results. One Verifier focuses on the ``related'' part, ensuring that all functions classified as related are indeed associated with computations or fund transfers. The other Verifier focuses on the ``unrelated'' part to prevent mistakenly discarding functions that may have implicit relevance. After the verification is completed, the two Verifiers send their respective results to the Combiner. The Combiner integrates the feedback from both Verifiers and makes final adjustments to the classification results to ensure the accuracy of pruning. The pruning results are subsequently passed to the symbolic execution detection tool for further defect analysis.

The symbolic execution detection tool contains four main components~\cite{yang2023definition}:the \textit{Inputter}, \textit{Feature Detector}, \textit{CFG Builder}, and \textit{Defect Identifier}. The \textit{Inputter} accepts user-provided Solidity source code and GPT's pruning results as input. It compiles the source code using various versions of Solidity compiler to obtain the bytecode and AST, and utilizes the API provided by Geth~\cite{geth} to disassemble bytecode into opcodes. The AST is analyzed to extract source mappings~\cite{sourcemap} for further analysis by other components.
The \textit{CFG Builder} performs symbolic execution and dynamically constructs the CFG while skipping the pruned function paths. It records key events (i.e., stack events, memory events, and call events) to detect defect features. During the CFG construction, the \textit{Feature Detector} identifies feature operations and maintains required data structures for detection (i.e., expression information, conditional comparisons, token flows, and internal\&external function calls). For each defect, the \textit{Defect Identifier} provides the final detection results based on the predefined features with the help of satisfiability modulo theories (SMT) solver~\cite{de2008z3}. 
%Feature Detector applies these rules to solve for the satisfiability of the defects. 
%Feature Detector applies these rules, passes the path conditions reaching the detection location and the constraint conditions of the defect rules to the satisfiability modulo theories (SMT)~\cite{de2008z3} solver for the solution, so as to determine whether the defect exists.
%Finally, the Defect Identifier confirms and reports the detected defects.

\subsection{Functions Pruning with GPT}

\noindent{\bf \textit{Motivations for Pruning}.} Traditional symbolic execution methods suffer from the problem of path explosion. To mitigate this problem and speed up the analysis process, we utilize LLMs for pruning by discarding entry function paths that are unrelated to numerical or transfer operations in the contract. It allows the symbolic execution framework to reach target functions more quickly.

Existing pattern-based pruning methods or manually designed heuristics struggle to handle scenarios involving complex semantics and intricate call relationships~\cite{he2021learning, wang2024contracttinker, ding2025smartguard}. Take transfer operations as an example, the implementation of transfer functions in smart contracts is diverse, using different variables for accounting and may be hidden in multi-level function calls, executed conditionally based on complex logic. These complexities make it difficult for pattern-based methods to accurately capture the semantics of such operations. 
In contrast, as stated in Section \ref{LLMscapabilities}, multiple studies have demonstrated that LLMs possess powerful code understanding and reasoning capabilities and can perform outstandingly in various software engineering tasks across multiple fields~\cite{zheng2025towards}.
LLMs can recognize complex semantics, analyze intricate calls, and determine whether a function is related to numerical operations or transfer operations from the source code level. Therefore, we use LLMs to accomplish the pruning task. For the possible inaccuracy and randomness in LLMs responses, we employ a multi-role collaboration strategy~\cite{ma2024combining} to mitigate these issues. It assigns distinct roles to different LLMs, with each role focusing on a specific aspect of the analysis. They cross-verify the outputs of others to enhance overall accuracy and reliability.

\noindent{\bf \textit{The pruning process}.} Although LLMs are capable of understanding and analyzing complex semantics, their responses may exhibit hallucinations~\cite{huang2023survey} and uncertainty~\cite{ouyang2023llm}. A multi-role collaboration strategy can reduce errors and randomness in the output~\cite{ma2024combining} and improve the effectiveness of pruning. The pruning process involves four roles: a Classifier, two Verifiers, and a Combiner. Considering both performance and cost, GPT-4o~\cite{gpt-4o} is selected as the large language model to implement these roles. As shown in Figure \ref{fig:prompt}, to further enhance the reliability of the answers and minimize the impact of randomness in GPT's output, we adopt the ``mimic-in-the-background'' prompting method, which is inspired by Sun et al.~\cite{sun2024gptscan}. With this method, GPT is prompted to simulate answering the same question five times in the background. The most frequently occurring answer is then selected to ensure higher consistency.

% \begin{figure}[b]
% \centering
%  \includegraphics[width=\linewidth]{figures/system.pdf}
%  % \centerline{System prompt.}
 
% \begin{minipage}{0.32\linewidth}
%     \vspace{3pt}
%     \centerline{\includegraphics[width=\textwidth]{figures/classifier.pdf}}
%     % \centerline{Image}
% \end{minipage}
% \begin{minipage}{0.32\linewidth}
%     \vspace{3pt}
%     \centerline{\includegraphics[width=\textwidth]{figures/verifier.pdf}}
%     % \centerline{Image}
% \end{minipage}
% \begin{minipage}{0.32\linewidth}
%     \vspace{3pt}
%     \centerline{\includegraphics[width=\textwidth]{figures/combiner.pdf}}
%     % \centerline{Image}
% \end{minipage}
% \caption{The Prompt Template Used by Roles.}
% \label{fig:prompt}
% \end{figure}

\begin{figure*}[t]
% \vspace{-1em}
    \centering
    \includegraphics[width=\linewidth]{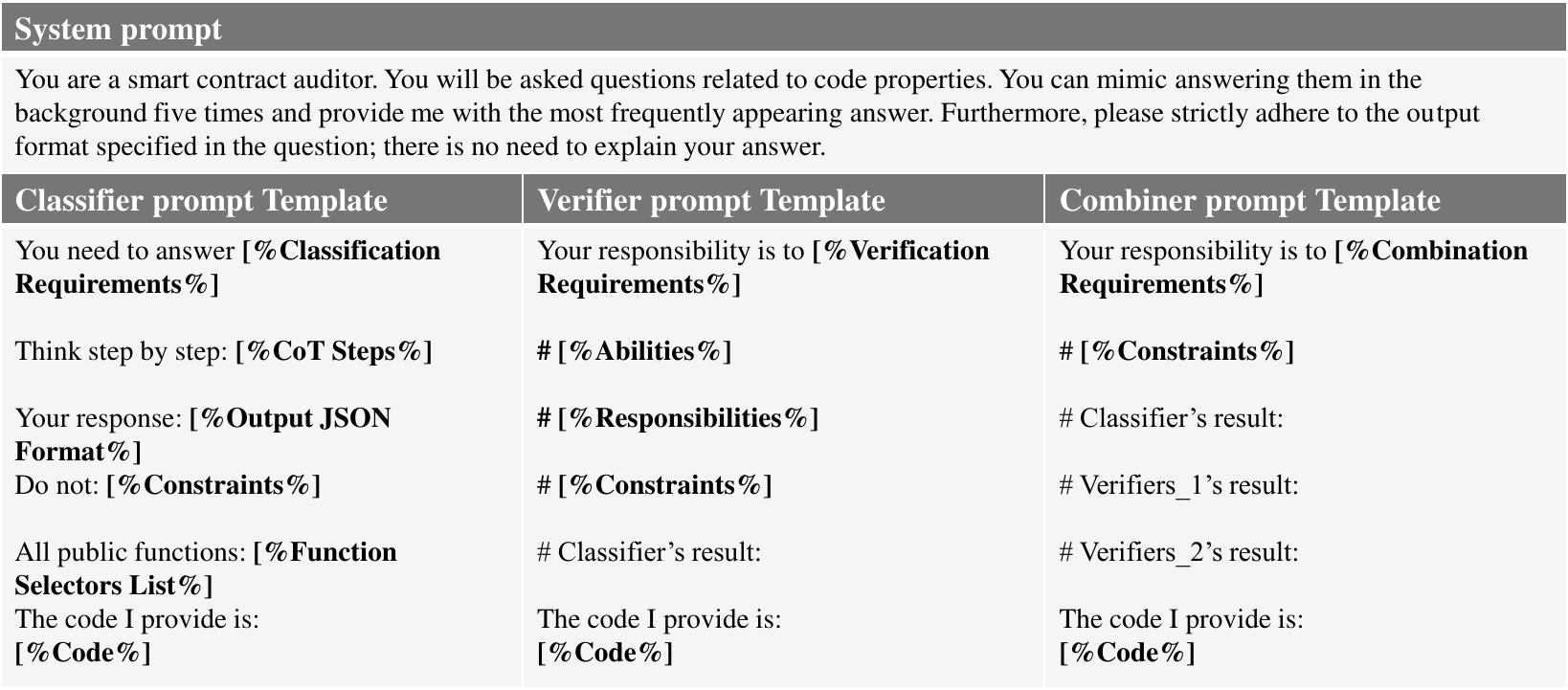}
    % \vspace{-1em}
    \caption{The Prompt Template Used by Roles.}
    \label{fig:prompt}
% \vspace{-1em}
\end{figure*}

\subsubsection{Rough Classification by Classifier} The Classifier is responsible for preliminary analysis and classification of function selectors, with the prompt template shown in Figure \ref{fig:prompt}. The prompt follows the zero-shot chain-of-thought (CoT)~\cite{kojima2022large} method. CoT guides LLMs to break down complex tasks into step-by-step logical reasoning sequences when answering questions, which enhances their reasoning capabilities and improves response reliability. [\%Classification Requirements\%] specify that the Classifier's task is to roughly categorize function selectors based on whether they are related to token numerical operations or balance changes, dividing them into two categories: ``related'' and ``unrelated''. The function selector is a 4 byte id that Solidity uses to identify functions. [\%CoT Steps\%] provides specific reasoning steps to guide GPT's inference process. First, it identifies which functions perform numerical operations on token or ether amounts. Then, it examines the relationships between function calls to determine which public functions from [\%Function Selectors list\%] call those numerical operation functions. Finally, it classifies them as either ``related'' or ``unrelated''. [\%Constraints\%] section highlights errors that GPT must avoid, e.g., ``Do not add functions that are not in the list''.

\subsubsection{Secondary verification by the Verifiers} Verifier\_1 is responsible for verifying ``related'' part of the classification results from Classifier, while Verifier\_2 is responsible for ``unrelated'' part. The results are categorized into two categories: ``correct'' and ``wrong''. The prompt templates used by both are shown in Figure \ref{fig:prompt}. Following the approach in ~\cite{ma2024combining}, we define more detailed capabilities, responsibilities, and constraints for GPT. [\%Abilities\%] require GPT to act as an excellent smart contract code reviewer, familiar with function calls. [\%Responsibilities\%] describe the task of examining the results from the Classifier and making judgments based on those results. [\%Constraints\%] define the required output format and certain types of errors that are not permissible.

\subsubsection{Merging Results by Combiner} Combiner synthesizes the reports from other partners to derive the final pruning results. [\%Constraints\%] emphasize that Combiner cannot simply merge the results but must exercise its own judgment. Additionally, it is essential for GPT to focus on the functions that highlight contradictions between the Classifier and the Verifiers when making the final determination.

\subsection{Symbolic Execution Detection Tool}

\subsubsection{Operational Semantics Modeling}
In this subsection, we model the syntax of several basic instructions, variables, and functions that form the foundation of the core analysis module for our detection tool.

We first present the operational semantics of two instructions related to function calls as follows. $CALL(f,o,l)$: Calls the target function $f$, with parameters loaded from memory starting at address $o$ with length $l$. $JUMPI(c,t)$: If the jump condition $c$ is satisfied, it jumps to location $t$ in the program.

During symbolic execution, certain key data structures are updated according to each executed EVM instruction. Specifically, $S$ represents the operand stack, $M$ denotes the simulated temporary memory space, and $GS$ represents the storage values of state variables. During path exploration, the constraints accumulated in the SMT solver are defined as the variable $cons$.
Our method utilizes source-level information provided by source mappings to assist in identifying defective code locations during symbolic execution. %Source mapping helps us find the function-level context.

\noindent{\bfseries Expression Information Recovery.} During the execution of the symbolic execution framework, expressions are simplified, resulting in the loss of expression information, i.e., operands and operators. Therefore, it is necessary to maintain a structure $v$ to store the information before the expression simplification. To utilize the expression information, we define a function $e(a, v)$ to recursively retrieve and recover information of the given expression $a$ from structure $v$, represented as $x, op, y := e(a, v)$. This allows us to analyze the original logical structure and meaning of expressions during symbolic execution.

\noindent{\bfseries Comparison Semantic Recognition.} In Solidity, there are three types of conditional statements: $if$, $require$, and $assert$. The $if$ statement enters the $else$ branch when the condition is not satisfied, whereas both $require$ and $assert$ revert the transaction and throw an exception. Compared to $require$ and $assert$, which have rollback protection and thus minimize the impact on the user in case of errors, $if$ statements pose a greater risk as they alter the execution path. Therefore, when analyzing patterns involving division within paths, the primary focus is on $if$ statements, making the identification of $if$ statements crucial.

The key to recognizing $if$ statements lies in identifying the comparison operator and retaining the two elements being compared. The comparison operator recognition is accomplished by matching source code with opcode sequences. Before executing the opcodes in a basic block, we first do a match process. If a corresponding opcode sequence and source code are matched, a trigger is set, and the relevant comparison operator is recorded. When executing the corresponding comparison opcode, e.g., $GT$ or $LT$, the two top values on the stack are captured. At this point, the two expressions being compared are obtained, which are used for subsequent detection.
We define function $cs(S)$ to retrieve the comparison elements $x$, $y$, and the comparison operator $cop$. The $seq$ represents the opcode sequence of comparison operation. Once these comparison elements are identified, we can further analyze the logic behind the conditional statements and track the execution flow during symbolic execution. This is crucial for defect detection, especially in scenarios where conditional logic may lead to different execution paths.

% \vspace{-0.5em}
$$x,cop,y=cs(S), \frac{S:=<x,y>,cop:=match(seq)}{GT|LT,JUMPI(jc,*)}$$

\begin{figure*}[t]
\begin{equation}
amount=ea(S, M),\left\{
\begin{aligned}
& \frac{S:=<*,*,amount,*,*,*,*>}{ctx(transfer), CALL(f,*,*)} \\
& \frac{amount:=M[o+x], x>4\And x<l}{ctx(transfer), CALL(f,o,l), S:=<*,*,*,o,l,*,*>} \\
& \frac{S:=<t,amount,arg2,...>}{ctx(transfer), JUMP(t)}
\end{aligned}
\right.
\label{amount-eq}
\end{equation}
\end{figure*}

\noindent{\bfseries External \& Internal Function Calls.} We focus on two types of transfer function calls: one where the token contract calls its own transfer function and another where it calls an external token contract's transfer function. For the former, the key opcode is $JUMP$, with parameters obtained from the stack. For example, when calling its own transfer function {\normalsize \texttt{transferFrom(from, to, amount)}}, the elements on the stack are arranged from top to bottom as follows: jump destination, amount, to and from.

For the latter, the key opcode is $CALL$, which first needs to retrieve the parameters' memory locations from the stack and then access the parameters from memory. For example, when calling an external token contract’s transfer function {\normalsize \texttt{token.transfer(to, amount)}}, the seven elements on the stack, from top to bottom, are: gas consumed, token contract address, ether amount to transfer, starting memory position of parameters ($o$), length of parameters($l$), starting memory position and length for storing the return data. Our tool reads the memory values to extract parameters based on the fourth and fifth elements. Specifically, the first four bytes at the starting memory position represent the function selector, i.e., $memory[o:o+4]$, followed by 32 bytes for $amount$, and another 32 bytes for $to$.

Recognizing internal and external calls is crucial for NumScout to acquire token transfer parameters, which is essential for detecting defects related to transfer amounts. Thus, we define the function $ea$ in Eq. \ref{amount-eq} to retrieve the expression for the transfer amount from stack and memory.

% \vspace{1em}
% \resizebox{\columnwidth}{!}{
% $$
% amount=ea(S, M),\left\{
% \begin{aligned}
% & \frac{S:=<*,*,amount,*,*,*,*>}{ctx(transfer), CALL(f,*,*)} \\
% & \frac{amount:=M[o+x], x>4\And x<l}{ctx(transfer), CALL(f,o,l), S:=<*,*,*,o,l,*,*>} \\
% & \frac{S:=<t,amount,arg2,...>}{ctx(transfer), JUMP(t)}
% \end{aligned}
% \right.
% \label{amounteq}
% $$
% }
% \vspace{1em}

\begin{figure*}[b]
\begin{equation}
\frac{T:=E(x)\cup E(y),\exists s\in T,s\in Input,i(a,b)}
{ctx(if),amount=ea(S,M),(x,>|\leq,y):=cs(S),(a,\div,b)=e(amount,v)}
\label{divinpath-eq}
\end{equation}
\end{figure*}

\subsubsection{Defects Detection}
To detect five new types of numerical defects in smart contracts, NumScout utilizes a symbolic execution framework to explore contract paths. A predefined semantic model assists in identifying execution states to capture key features and locate defects. In the following paragraphs, we provide a detailed explanation of how these defects are discovered in smart contracts.

\noindent {\bfseries (1) Div In Path:} The tool first needs to locate the $if$ comparison statements. Using the opcode sequence matching method mentioned above, it applies function $cs$ to extract the comparison operator and two elements.
The presence of this defect requires three conditions to be satisfied. First, the division operation must be on the left side of $>$ operator, and if subtraction is involved, the sides must be switched. Second, the division must be indivisible. Third, the two expressions being compared must contain user input values, meaning that the user input can influence the comparison result and thus affect the program's execution path.
The reason why the first condition is necessary is that if the division operation appears on the left side of $<$, e.g., \verb|if(a/100<3)|, then users may believe that the condition is satisfied when $a<3*100$, which is indeed correct. The case where the division operation is on the right side of $>$ is discussed in Section \ref{ifGT}, where inexperienced users are more likely to be misled. The $\leq$ operator can be seen as an alternative branch of $>$, and both can be viewed as the same situation; similarly, $\geq$ and $<$ represent another equivalent situation.

Function $E$ is used to extract the symbolic variables within the expressions. Function $i(a, b)$ checks whether $a$ can divide $b$ without leaving a remainder, which could lead to precision loss. The SMT solver adds this condition to the constraints and solves for the satisfiability. The detection rule for this defect is shown in Eq. \ref{divinpath-eq}.

\noindent {\bfseries (2) Operator Order Issue:} The defect of computation order requires recovering the information of the expression $amount$ from $e(amount, v)$ to construct an operator order tree, which is accomplished by function $bt$. This function ensures that all operators are organized in a hierarchical structure, preserving their precedence as defined in the original expression. Then, a depth-first traversal of the tree is performed. This step is carried out by the $dfs$ function, which takes the operator order tree returned by $bt$ as input. It begins traversal from the root node and explores each node following a depth-first strategy, inspecting the operator types along the way. If the pattern that division occurs before multiplication appears along any path, it indicates the presence of \textit{Operator Order Issue}.

% \vspace{-0.5em}
$$
\frac{tree:=bt(e(amount,v)),dfs(tree)}
{ctx(transfer),amount=ea(S,M)}
$$

\noindent {\bfseries (3) Minor Amount Retention:} The defect of \textit{Minor Amount Retention} requires not only that the transfer amount expression presents the possibility of being indivisible, but also that no other path exists to transfer the total ether or tokens held by the contract; otherwise, the retention defect does not exist. The expressions below illustrate our detection logic for \textit{Minor Amount Retention}, where all possible paths for transferring ether and tokens are defined as $path$, and the path for transferring all ether or tokens is denoted as $P$.

% \vspace{-0.5em}
$$
\frac{(x,\div,y):=e(amount,v),i(x,y),\nexists P\in path}
{ctx(transfer),amount=ea(S,M)}
$$

\noindent {\bfseries (4) Exchange Problem:} Our tool records the token flow in a structure $t$ during the symbolic execution if it identifies a transfer operation. This is used to detect token exchange defects. Each flow consists of three elements: $from$, $to$, and $amount$. The function $f$ is defined to find the two types of tokens involved in the exchange and their corresponding exchange amounts from structure $t$. From the contract’s perspective, $in=0\wedge out\neq0$ indicates that the user may gain profit for free. Conversely, the scenario $in\neq0\wedge out=0$ indicates that rounding errors may occur during the exchange. Both conditions are passed to the SMT solver for checking. If the result is satisfiable, it can be confirmed that the \textit{Exchange Problem} defect exists.
% the presence of \textit{Exchange Problem} is confirmed.

% \vspace{-0.5em}
$$
\frac{(in,out):=f(t),in=0\wedge out\neq0 | in\neq0\wedge out=0}
{ctx(transfer)}
$$

\noindent {\bfseries (5) Precision Loss Trend:} To detect the defect of \textit{Precision Loss Trend}, the function $rt$ is first used to parse the expression and determine the rounding method. Specifically, it analyzes the operands and operators restored from the expression. If the numerator has been incremented by (denominator-1) before the division operation, the rounding type is identified as rounding up, i.e., $ceil(x)$. Then, the token flow’s $from$, $to$, and rounding method are analyzed together using function $ct$. If the rounding method is $ceil(x)$ and the flow is outgoing from the contract, or if there are two or more flows with the same $from$ but different $to$ and different rounding methods, the contract contains this defect. The former indicates that the direction of precision loss does not meet the requirements for maintaining the liquidity pool, while the latter suggests that there is unfair reward distribution.

% \vspace{-0.5em}
$$
\frac{r:=rt(e(amount,v)),ct(from,to,r)}
{ctx(transfer),amount=ea(S,M)}
$$

\section{Experiment} \label{section:experiment}

In this section, based on an open-source dataset, we first conduct a small-scale experiment and evaluate the effectiveness of NumScout. We also perform detection on the large-scale dataset to confirm the situation of numerical defects in real-world contracts. Ablation experiments are conducted to demonstrate the effectiveness of the GPT-based pruning component.

\subsection{Experimental Setup}

The experiment is conducted on a server running Ubuntu 22.04.2 LTS, with a configuration of 20 Intel Xeon Platinum 8360H CPUs and 200GB of memory.

\noindent {\bfseries Dataset.} To determine the prevalence of the defined defects in real-world Ethereum smart contracts, we utilize an open-source dataset from a GitHub repository~\cite{smartcontractsanctuary}, which stores the source code of all verified smart contracts on Etherscan up to July 13th, 2023. We downloaded this dataset on September 20th, 2024, and selected contract files deployed on the Ethereum mainnet, totaling 331,382 mainnet contracts. The dataset provides a summary file with basic information about each contract, e.g., contract's address, ether balance, compiler version, and total number of transactions. To filter for valuable contracts, we apply two criteria: total transactions $>$ 100 and ether balance $>$ 0. We further classify contracts by compiler version and remove those that cannot be compiled. These two filtering conditions ensure that the selected contracts are actively used by users in real-world scenarios, rather than toy contracts. This selection allows the experimental results to better reflect the tool’s performance in detecting defects in widely used real-world contracts. Ultimately, we obtain \LargeDatasetNum~contracts.

%Our dataset more accurately captures the characteristics of current Solidity smart contracts, which exhibit much more complexity compared to those with earlier versions. 
We compare our dataset with SmartBugs~\cite{ferreira2020smartbugs}, a widely used dataset. Table \ref{table:DatasetCompare} highlights key characteristics of both datasets. It is evident that our dataset contains more complex smart contracts than those in SmartBugs. Specifically, the average lines of code (LOC) and the number of instructions of contracts in our dataset are \textbf{11.5X} and \textbf{5.5X} higher compared to SmartBugs, respectively. Additionally, the average number of public/external functions and state variables in our dataset are approximately \textbf{3X} and \textbf{4X} higher than in SmartBugs. \veightperct~of the contracts in our dataset require a Solidity compiler version higher than v0.8.0, whereas 99.4\% of the contracts in SmartBugs rely on versions below v0.5.0.

\begin{table}[h]
\centering
  \caption{Features of Our Dataset vs. SmartBugs.}
\resizebox{\columnwidth}{!}{
  \begin{tabular}{c|c|c|c|c}
    \hline
    \diagbox{\textbf{Dataset}}{\textbf{Features}} & \textbf{LOC} & \textbf{\#of Instrs} & \textbf{\#of Funs} & \textbf{\#of State Vars} \\
    \hline
    \hline
    Ours & \LOC~& \Instrs~& \Funs~& \StateVars~\\
    \hline
    SmartBugs & 99.9 & 1545.5 & 12.5 & 6.6 \\
    \hline
  \end{tabular}}
% \vspace{-1em}
\label{table:DatasetCompare}
\end{table}

\noindent {\bfseries  Evaluation Metrics.} We outline the following research questions (RQs) to assess the effectiveness of NumScout.

\begin{itemize}
    \item \textbf{RQ1:} What is the efficacy of NumScout in detecting the five new types of defined numerical defects? 
    \item \textbf{RQ2:} How effective is NumScout in detecting defects within our large-scale dataset?
    \item \textbf{RQ3:} How effective is the pruning component based on GPT?
\end{itemize}

\subsection{Answer to RQ1: Evaluation of NumScout}

\begin{table*}[t]
\centering
% \vspace{-1em}
\normalsize
  \caption{Defects in Samples and Evaluation of NumScout.}
  \begin{tabular}{c|c|c|c|c|c|c|c}
    \hline
    \textbf{Defect} & \textbf{all} & \textbf{TP} & \textbf{FP} & \textbf{FN} & \textbf{Precision(\%)} & \textbf{Recall(\%)} & \textbf{F1-score(\%)} \\
    \hline
    \hline
    \textit{Div In Path} & 7 & 7 & 0 & 1 & 100.0 & 87.5 & 93.3 \\
    \hline
    \textit{Operator Order Issue} & 7 & 7 & 0 & 3 & 100.0 & 70.0 & 82.4 \\
    \hline
    \textit{Minor Amount Retention} & 19 & 15 & 4 & 5 & 78.9 & 75.0 & 76.9 \\
    \hline
    \textit{Exchange Problem} & 3 & 3 & 0 & 1 & 100.0 & 75.0 & 85.7 \\
    \hline
    \textit{Precision Loss Trend} & 3 & 3 & 0 & 1 & 100.0 & 75.0 & 85.7 \\
    \hline
  \end{tabular}
  % \vspace{-1em}
\label{table:samplesRes}
\end{table*}

To answer RQ1, we randomly sample a subset from the large-scale dataset for a small-scale experiment, where all samples are checked and labeled manually. Specifically, to determine the sample size, we follow a sampling method based on confidence intervals~\cite{confidenceinterval} to generalize the detection results from the sample to the overall dataset. We set a 10 confidence interval and 95\% confidence level, calculating the required sample size to be 95. We randomly select a sample dataset and run NumScout on it. Two of the authors manually label the results of all samples carefully. We first collaboratively discuss and label 30\% of the sample results to establish and confirm the labeling criteria. Then, we independently label the remaining 70\% of the results, followed by a comparison and integration of the final results. We separate true positives (TP), false positives (FP), true negative(TN) and false negative (FN) during the labeling process to analyze the performance of NumScout. This method is also employed in other related works~\cite{luu2016making,jiang2018contractfuzzer,kalra2018zeus}.

Table \ref{table:samplesRes} displays the performance of NumScout on the labeled samples.
The fifth to seventh columns show the number of TP, FP, FN in the samples, respectively. We use Precision $P=\frac{TP}{TP+FP}$, Recall $R=\frac{TP}{TP+FN}$, and F1-score $F1=\frac{2*P*R}{P+R}$ to measure the detection performance for each type of defect. Additionally, we calculate the overall precision to demonstrate the effectiveness of NumScout. It can be calculated as $\frac{\sum_{i=1}^n p_{c_i} \times\vert c_i \vert}{\sum_{i=1}^n \vert c_i \vert}$, where $p_{c_i}$ represents the precision of detecting defect $i$, and $\vert c_i \vert$ is the number of defect $i$ in our dataset. 
NumScout achieves 100\% precision in detecting the \textit{Div In Path}, \textit{Operator Order Issue}, \textit{Exchange Problem} and \textit{Precision Loss Trend}. For the \textit{Minor Amount Retention}, it reports them at 78.9\% precision. Overall, the comprehensive precision reaches \OverallPrecision.

{\bfseries False Positives.} Our experimental results contain some false positives in the detection of \textit{Minor Amount Retention} due to an inability to recognize specific transfer paths.
In certain contracts, users input a percentage number to withdraw funds from the contract as Figure \ref{fig:FPsample}. The contract provides a $clearStuckBalance$ function, where the input parameter $amountPercentage$ represents the percentage of the total balance that the owner intends to withdraw. The function first ensures that the input $amountPercentage$ does not exceed 100\%. It then calculates the withdrawal amount by multiplying the contract’s total balance $amountBNB$ by the input percentage and dividing by 100. Finally, it transfers the computed amount to the designated wallet $\_marketingWallet$. Users can withdraw funds at a 100\% ratio, which creates a path that allows all funds to be transferred out. However, our tool fails to identify this special path for proportional fund withdrawal. Instead, due to its inability to track and analyze dynamic fund withdrawal conditions, it interprets the contract as potentially retaining a minor amount of funds. Consequently, this limitation leads to false positives when identifying the Minor Amount Retention defect.

\begin{figure}[h]
\centering
% \vspace{-1em}
\begin{lstlisting}
function clearStuckBalance(uint256 amountPercentage) external onlyOwner {
    require(amountPercentage <= 100);
    uint256 amountBNB = address(this).balance;
    payable(_marketingWallet).transfer(amountBNB.mul(amountPercentage).div(100));}
\end{lstlisting}
% \vspace{-0.5em}
\caption{A FP case of Minor Amount Retention defect.}
\label{fig:FPsample}
\end{figure}

{\bfseries False Negatives.} We find that among the 95 samples, 11 are false negatives. All of the missed reports are caused by path explosion. Specifically, these contracts contain multiple branches in their CFG, leading to a huge search space. To avoid path explosion, we limit the tool's maximum loop iterations, the depth of path exploration, and the execution time. As a result, NumScout fails to detect the locations of these defects. 
It is worth noting that the main purpose of GPT-based pruning is to discard unrelated entry function paths, allowing the symbolic execution framework to reach the target function more quickly. However, it does not address the issue of overly deep search paths within the function.

For example, in one of the missed contracts\footnote{\url{aaf740FD71093520C457642eb9219A4F6dA22190}}, there are seven $require$ statements and five $if$ conditional statements (lines 903-929) preceding the defective code, making the search paths extremely complex and causing NumScout to miss the defect as a consequence.

To mitigate false negatives caused by path explosion, the following optimization strategies can be considered. One approach is to introduce a heuristic search strategy. LLMs rank all functions based on their relevance to numerical operations and the risk level of fund transfers, prioritizing the exploration of paths more likely to contain defects. LLMs can also integrate with dynamic symbolic execution to intelligently adjust subsequent search directions based on previously explored paths. Additionally, preprocessing complex control flow structures helps simplify the search space by flattening excessively nested loops and conditional branches where appropriate.

\subsection{Answer to RQ2: Defects Detection in a Large-Scale Dataset}

To address RQ2, we run NumScout on the source code of all the collected \LargeDatasetNum~verified smart contracts, which includes the 95 samples in RQ1. Table \ref{table:largeRes} provides the numbers and frequency of each new type of numerical defect in contracts on Ethereum. NumScout only identifies whether a defect exists in the contract, so if the same type of defect appears multiple times, we count it only once.

\textit{Minor Amount Retention} is the most common defect in our dataset, present in approximately 15.1\% of the smart contracts. About 8.6\%, 4.7\%, and 1.7\% of contracts contain the \textit{Div In Path}, \textit{Minor Amount Retention}, and \textit{Precision Loss Trend} defects, respectively. Moreover, the proportions of \textit{Exchange Problem} are all below 1\%, with 39 (0.60\%) smart contracts containing this defect.

Additionally, the experimental results indicate that 45 smart contracts contain 3 types of the 5 defined defects, while 194 smart contracts include 2 types of defects. Overall, as reported by NumScout, there are \LargeDefectNum~smart contracts that have at least one type of defect in our dataset, which accounts for 26.8\% of all contracts.

\begin{table}[t]
% \vspace{-1em}
\centering
\normalsize
  \caption{Defects in Large-Scale Dataset.}
  % \vspace{-5pt}
  \begin{tabular}{c|c|c}
    \hline
    \textbf{Defect} & \textbf{\# Defects} & \textbf{Percentage(\%)} \\
    \hline
    \hline
    \textit{Div In Path} & 561 & 8.6 \\
    \hline
    \textit{Operator Order Issue} & 306 & 4.7 \\
    \hline
    \textit{Minor Amount Retention} & 983 & 15.1 \\
    \hline
    \textit{Exchange Problem} & 39 & 0.60 \\
    \hline
    \textit{Precision Loss Trend} & 114 & 1.7 \\
    \hline
  \end{tabular}
  % \vspace{-1em}
\label{table:largeRes}
\end{table}

{\bfseries Contracts with \textit{Minor Amount Retention} Defects.} The large-scale experiment reveals that the number of contracts containing \textit{Minor Amount Retention} defects is significantly higher than that of other defects. We find that many projects encounter cases where the profits cannot be divided evenly during distribution. Although, in the long term, the retained amount is only a small portion, represented as a random variable parameterized by the number of players, we consider that these small retained balances might be referenced by other contracts. Attackers may exploit this situation to inflict potentially substantial losses on other contracts.

\subsection{Answer to RQ3: Ablation Experiment Results}
In RQ3, we evaluate the effectiveness of the GPT-based pruning component. Specifically, we conduct an ablation experiment on the selected samples by removing the GPT-based pruning component. In this setup, the tool does not receive the list of functions unrelated to numerical operations or transfers, forcing it to explore all execution paths. The results show that the tool with pruning runs \improvement~faster than the version without pruning and identifies two additional \textit{Operator Order Issue} defects. Specifically, to detect more defects, we set a time limit of 1,800 seconds and a search depth limit of 500 during the experiment. Additionally, we allow a longer SMT solver satisfiability checking time of 600 seconds at critical verification points. For each contract, the average runtime of the tool without the GPT-based pruning component reaches 1,518.56 seconds, while incorporating the GPT-based pruning component reduces the average runtime to 1,182.19 seconds. The average cost of the entire pruning process for a single contract on GPT-4o is only \$0.008. Figure \ref{fig:ablationeg} illustrates a defect detected in the RQ1 experiment but missed during the ablation experiment. The tool must execute the $sell$ function first, following specific paths to modify certain variables before the defect condition at line 6 is satisfied. The $sell$ function contains 9 function calls, 7 conditional statements, and about 40 numerical operations, resulting in a huge search space for symbolic execution.

Notably, our tool not only identifies the code location where the defect is triggered but also provides the entire call path, helping developers trace the defect’s origin. From the results, we observe that the pruned version of the tool reaches the defect trigger point twice through different paths within the time limit. In contrast, without pruning, the tool wastes execution time in other unrelated functions, preventing symbolic execution from reaching the critical path within the time limit.
The ablation experiment confirms that pruning enables the tool to enter target functions more quickly, improving detection speed and identifying more defects.

\begin{figure}[H]
\centering
% \vspace{-1em}
\begin{lstlisting}
function exit() public {
    if(_tokens > 0) sell(_tokens);
    withdraw();}
function withdraw() onlyStronghands() public {
    uint256 _dividends = myDividends(false);
    _customerAddress.transfer(_dividends);}
\end{lstlisting}
% \vspace{-0.5em}
\caption{A case of undetected Operator Order Issue defect in the ablation experiment.}
\label{fig:ablationeg}
% \vspace{-0.5em}
\end{figure}

\section{Discussion} \label{section:discussion}

\subsection{Case Study}

We present a real-world case\footnote{\url{0x3c07b3f4a6e253915d83c86707f0af07521d1cd8}} from the tool's report to illustrate how a user loses funds due to the new types of numerical defect and demonstrate the importance of detecting these defects reported by NumScout. Figure \ref{fig:casestudy} displays a simplified code snippet from the affected contract.

%The user can purchase tokens by sending ether when calling the $sale$ function, but the token amount is calculated using a divide-then-multiply order. Given that the value of $cloudsPerEth$ on the current blockchain is 800,000, inexperienced users may assume that 1,250,000,000 wei is sufficient to buy 1 token. In reality, if users send less than 0.001 eth, the integer division results in $amount = 0$. Since the contract does not check the token amount exchanged, the transaction does not revert but continues to execute, leaving the users without any received tokens and causing them to lose the ether they sent. Our tool identifies this defect from two aspects: \textit{Operator Order Issue} and \textit{Exchange Problem}. The former is detected by analyzing the expression operator tree, while the latter is detected through token flow analysis.

The user can purchase tokens by sending ether when calling the $sale$ function, but the token amount is calculated using a divide-then-multiply order. Given that the value of $cloudsPerEth$ on the current blockchain is 800,000, inexperienced users may assume that 1,000,000,000,000,000 wei (i.e., 0.001 ether) is equivalent to 800,000 tokens, which means 1,250,000,000 wei is sufficient to buy 1 token. However, if the user sends less than 0.001 ether, the integer division results in $amount = 0$. Since the contract does not check the token amount exchanged, the transaction does not revert but continues to execute, leaving the users without any received tokens and causing them to lose the ether they sent. Our tool identifies this defect from two aspects: \textit{Operator Order Issue} and \textit{Exchange Problem}. The former is detected by analyzing the expression operator tree, while the latter is detected through token flow analysis.

\begin{figure}[h]
\centering
% \vspace{-1em}
\begin{lstlisting}
function sale() payable {
    uint256 amount = (msg.value / 1000000000000000) * cloudsPerEth;
    balances[msg.sender] += amount;
    balances[owner] -= amount;
    Transfer(owner, msg.sender, amount);}
\end{lstlisting}
\caption{Code snippet of the Operator Order Issue and Exchange Problem case.}
% \vspace{-1em}
\label{fig:casestudy}
\end{figure}

We verify the funds loss process on the local test network with two test accounts. The first account deploys the contract, being the holder of the total token supply (i.e., $owner$), sets $cloudsPerEth$ to match its current value on the mainnet, and enables the trading switch. The second account (i.e., $msg.sender$) calls the $sale$ function, sending 0.0005 ether and expecting to receive 400,000 tokens. The result shows that while the ether balance of $msg.sender$ decreases, token balance remains 0. Meanwhile, the ether balance held by the contract increases by 0.0005 eth, and the $owner$'s token balance does not decrease. The verification script is available in our online repository.

\subsection{Implications}

\noindent {\bfseries For Researchers.} Blockchains based on the EVM and supporting smart contract development in Solidity may exhibit similar numerical defects, though they may exhibit different patterns due to variations in blockchain characteristics. This possibility enables researchers to conduct further analysis and suggests new directions for future research.

\noindent {\bfseries For Practitioners.} %We present five types of new numerical defects in contracts to help practitioners better understand their definitions and consequences. 
For developers, the defined defects aid in gaining a deeper understanding of numerical operations involved in smart contracts, particularly issues related to rounding and precision loss. It reminds developers to pay attention to minor precision losses and to improve testing efforts. These numerical defects can serve as coding guidance for developers during contract development to ensure robustness. For auditors, it enhances their awareness of the security about numerical operations, encouraging them to adopt more comprehensive auditing strategies.

\noindent {\bfseries For Investors and Users.} It is important for investors and users to be cautious about potential numerical defects in contracts, which are often hidden within complex mathematical operations and can be difficult to detect. Additionally, our tool can help identify losses that may arise from numerical defects and in flagging contracts that might exploit these defects for fraudulent purposes. %It reminds users to value audit reports and ensure that the contracts they invest in have undergone professional security audits.

\noindent {\bfseries For Educators.} In smart contract development courses, educators should provide best practices and share known cases for avoiding numerical defects. This helps students recognize these defects and the serious consequences they may cause.

\subsection{Threats to Validity}

\noindent {\bfseries Internal Threats.} One potential internal threat in our study is that we did not analyze all available audit reports, which may have led to the omission of some numerical defects. However, we mitigated this risk by utilizing an iterative information retrieval strategy to extract as many audit reports related to numerical defects as possible. The reports collected through this keyword-based approach help ensure comprehensive coverage and minimize the risk of missing relevant defects. Another internal threat arises from the high complexity of the smart contracts in our dataset, which makes the symbolic execution process highly time-consuming. Additionally, new types of numerical defects often involve division operations, which are computationally difficult for SMT solvers and require significant time to process. We address the execution time issue using GPT-based pruning, and ablation results confirm the effectiveness of this approach.

\noindent {\bfseries External Threats.} Our dataset is filter based on specified criteria, which may have excluded numerical defects present in other contracts. However, by filtering contracts with more than 100 transactions and non-zero balances, our dataset reflects the numerical defects found in frequently used real-world contracts rather than those in test or toy contracts, providing a better evaluation of our tool’s effectiveness. During the manual labeling process,  there may be instances of incorrectly classifying false negatives and true negatives. To address this, we adopt a double-check mechanism and update the labeled dataset in a timely manner to ensure accuracy.

\subsection{Possible Solutions for the five numerical defects} 
%To ensure the numerical safety of contracts, we not only design NumScout to detect potential defects but also aim to assist developers in developing secure smart contracts. Therefore, 
In this subsection, we provide recommendations for developers to avoid introducing the defined five types of numerical defects in contracts. Section \ref{section:defects} presents defect code examples from audit reports, along with suggestions provided by security teams. We summarize the recommended fixes from the remaining audit reports, listing brief solutions for each type of defect in Table \ref{table:solutions}. It is worth noting that in the \textit{Operator Order Issue} defect, when changing the code from division before multiplication to multiplication before division, it requires careful consideration of overflow risks. For example, in the defective code shown in Figure~\ref{fig:casestudy}, if line 2 is modified to {\normalsize \texttt{uint256 amount = msg.value * cloudsPerEth / 1000000000000000}}, it is important to note that $msg.value * cloudsPerEth$ might exceed the maximum value of $uint256$. As a result of the overflow, it will become a small number. This situation could occur when a user transfers a large amount of ETH to exchange for tokens, potentially causing losses of user funds. If the contract uses Solidity version v0.8.x, the compiler will automatically insert overflow checks into the bytecode, eliminating the need for developers to handle overflow risks. For versions lower than v0.8.0, developers should use the SafeMath library to prevent potential overflow.

\begin{table}[h]
  \centering
  \normalsize
  \setlength\tabcolsep{1.5pt}
  \caption{Possible Solutions for the five Defects.}
\resizebox{\columnwidth}{!}{
  \begin{tabular}{>{\centering\arraybackslash}m{0.35\linewidth} | >{\arraybackslash}m{0.65\linewidth}}
    \hline
    \textbf{Defect} & \textbf{Possible Solution} \\
    \hline
    \hline
    \textit{Div In Path} & Use multiplication instead of division in conditional statements. \\
    \hline
    \textit{Operator Order Issue} & Multiply first and divide later, but be cautious of overflow. \\
    \hline
    \textit{Minor Amount Retention} & Implement a function to withdraw all funds. \\
    \hline
    \textit{Exchange Problem} & Check the calculation results before the transfer. \\
    \hline
    \textit{Precision Loss Trend} & Consider who bears the loss of precision. \\
    \hline
    \textit{General Advice} & Conduct thorough rounding tests. Avoid letting the liquidity pool bear precision loss. Ensure consistent precision between both tokens in the swap. \\
    \hline
  \end{tabular}}
  % \vspace{-1em}
\label{table:solutions}
\end{table}

Several additional suggestions for numerical operations are as follows: 
Developers need to thoroughly test rounding boundaries and rounding effects before deploying the contract.
If there is a precision rounding issue in trading pool or lending pool, it is best not to let the liquidity pool bear the precision loss. Instead, calculations should favor the liquidity pool to ensure the pool remains balanced.
Check the precision of the two assets being exchanged to prevent unexpected results due to differing precisions. For example, most ERC20 standard tokens have 18 decimals~\cite{erc20}, while tokens like USDT have only 6 decimals. Some contracts do not handle these situation, leading to security problem.
If calculations indeed involve assets with different precisions, prioritize using the asset with lower decimals for calculations. Then, derive the amount of the asset with higher decimals through multiplication. This way, all mathematical operations are based on multiplication, avoiding the creation of decimal units.
To validate the effectiveness of these solutions, we randomly select 10 contracts for each type of defect and apply the recommended fixes. We then analyze these revised contracts using NumScout, and the results show that NumScout reports no defects.

\section{Related Work} \label{section:relatedwork}

\subsection{Smart Contract Defects} Chen et al. propose the first research that defines smart contract defects from the developers' perspective~\cite{chen2020defining}. They collect posts from StackExchange and use an open card sorting method to discover and categorize 20 types of contract defects. Additionally, they design a survey to gather developers' feedback and concerns regarding these defects. In another work, they introduce a tool named DefectChecker~\cite{chen2021defectchecker}, which detects these defined defects by analyzing the bytecode of contracts.
However, their research do not cover the new types of numerical defect that arise in smart contracts. Specifically, they defines $Unmatched Type Assignment$, which focuses on mismatches between assignments and types, potentially leading to integer overflow. This differs from the new numerical defects we focus on, which can result in transaction execution errors.

\subsection{Tools for smart contract defects detection} Many program analysis tools focus on detecting traditional numerical defects. Luu et al. proposed the first symbolic execution-based tool, Oyente~\cite{luu2016making}, which simulates EVM instruction execution and explores different execution paths to construct CFG. It uses the Z3 SMT solver to determine whether vulnerability conditions are satisfied, enabling the detection of overflow vulnerabilities. Torres et al. introduced a framework called Osiris~\cite{torres2018osiris} that identifies three types of integer-related defects in Ethereum smart contracts through taint analysis: Arithmetic Bugs, Truncation Bugs, and Signedness Bugs. Additionally, other static analysis tools such as MAIAN~\cite{nikolic2018finding}, Securify~\cite{tsankov2018securify}, Ethainter~\cite{brent2020ethainter}, Sailfish~\cite{rao2012sailfish}, Mythril~\cite{mythril}, and Slither~\cite{feist2019slither} have also been developed to detect defects in Solidity smart contracts. Meanwhile, tools like ContractFuzzer~\cite{jiang2018contractfuzzer}, sFuzz~\cite{nguyen2020sfuzz}, Smartian~\cite{choi2021smartian}, and Echidna~\cite{grieco2020echidna} are based on dynamic testing and analysis.

%However, all of these tools focus solely on overflow or truncation errors, while the numerical defects discovered in this paper serve as a complementary contribution to these existing approaches.

\subsection{Accounting Errors} Another type of smart contract defect involving numerical operations is the Accounting Errors. They are specific to the financial logic of the contract, focusing on incorrect financial logic operations, such as adding fees to a user’s balance instead of deducting them or directly summing tokens of different units. The tool ScType~\cite{zhang2024towards} models financial operations and high-level information in DeFi, e.g., token units, scaling factors, and financial types, and leverages type propagation and checking to detect Accounting Errors. ScType relies on specific business contexts and requires manual completion of initial type annotations. In contrast, our work complements this work by focusing on issues arising from the nature of numerical calculations themselves, such as precision loss and improper operator order, which may lead to unexpected behaviors during smart contract execution.

\subsection{LLMs in Smart Contract Defect Detection} Currently, LLMs are widely used in smart contract defect detection. Sun et al. propose GPTScan~\cite{sun2024gptscan}, the first tool that integrates GPT with static analysis for detecting logical vulnerabilities in smart contracts. This tool uses GPT to identify key variables and statements, followed by static analysis to verify potential vulnerabilities. Ma et al. introduce the iAudit framework~\cite{ma2024combining}, which combines LLM fine-tuning with a multi-role strategy to audit contracts through iterative debates. Ding et al. propose SmartGuard~\cite{ding2025smartguard}, a framework that retrieves semantically similar code, generates CoT, and then utilizes LLMs for vulnerability identification. Wang et al. present ContractTinker~\cite{wang2024contracttinker}, which also employs CoT and program static analysis to guide LLMs in repairing real-world smart contract vulnerabilities. Wu et al. develop AdvSCanner method~\cite{wu2024advscanner}, which uses static analysis to extract attack flows related to reentrancy vulnerabilities and utilizes them to guide LLMs in generating attack contracts that can exploit reentrancy vulnerabilities in victim contracts. Our work uses LLMs for pruning and combines symbolic execution tools to confirm new types of numerical defects, which expands and complements these existing works.

\section{Conclusion}\label{section:conclusion} There are two main parts in this paper: the definition of defects and their detection. We summarize five new types of numerical defect patterns from the audit reports provided by the DAppScan dataset, which are collected from multiple renowned blockchain security teams. These issues are considered high-risk and impact the execution results of programs. For each defect, we provide code examples and possible solutions. To identify defects in real-world smart contracts, we develop a tool called NumScout, which utilizes GPT-based pruning and symbolic execution to detect the aforementioned five defined defects.

NumScout uses GPT-4o for pruning, removing functions unrelated to numerical operations and transfers, thus enhancing the efficiency of subsequent symbolic execution. The tool performs symbolic execution at the bytecode level, combined with source code features for analysis. Specifically, the tool constructs and analyzes expression operator order tree, extract the conditional statements of comparison from the bytecode, analyzes token flows, and other methods to extract key features. It reports defects based on predefined defect patterns combined with source code mapping. Moreover, NumScout supports all compiler versions and is extensible, allowing developers to write additional detection patterns to identify more defects. Experimental results show that NumScout identifies \LargeDefectNum~smart contracts containing at least one defined defect in the dataset. Furthermore, NumScout achieves an overall detection precision of \OverallPrecision.

\section*{Acknowledgment} This work is partially supported by the Zhejiang Provincial Key Project of Undergraduate Education and Teaching Reform (JGZD2024060), the Zhejiang Provincial Higher Education Research Project \& Special Research Project on Artificial Intelligence Empowering Education and Teaching Applications (KT2024007), the Sichuan Provincial Natural Science Foundation for Distinguished Young Scholars (2023NSFSC1963), and the National Natural Science Foundation of China (62332004).